\documentclass[aps,
               pra,
               twocolumn,
               amsmath,
               amssymb,
               superscriptaddress,
               reprint,
               10pt
              ]{revtex4-2}

\usepackage{url}
\usepackage{txfonts}

\usepackage{amsmath, amssymb}

\usepackage{pstricks}
\usepackage{graphicx}
\graphicspath{{./figures/}}
\usepackage{tikz}
\usepackage[T1]{fontenc}
\usepackage{color}
\usepackage[colorlinks=true,breaklinks=true,allcolors=blue]{hyperref}
\usepackage{tabularx}
\usepackage{multirow}
\usepackage{array}
\newcolumntype{L}[1]{>{\raggedright\let\newline\\\arraybackslash\hspace{0pt}}m{#1}}
\newcolumntype{C}[1]{>{\centering\let\newline\\\arraybackslash\hspace{0pt}}m{#1}}
\newcolumntype{R}[1]{>{\raggedleft\let\newline\\\arraybackslash\hspace{0pt}}m{#1}}

\usepackage{physics}
\usepackage{xparse}
\usepackage[version=4]{mhchem}
\usepackage{lipsum}
\usepackage[normalem]{ulem}

\renewcommand{\i}{\mathrm{i}}
\newcommand{\e}{\mathrm{e}}

\renewcommand{\vec}[1]{\mathbf{#1}}

\newcommand{\eq}[1]{(\ref{eq:#1})}
\newcommand{\Eq}[1]{Eq.\,\eqref{eq:#1}}

\newcommand{\Fig}[1]{Fig.~\ref{fig:#1}}
\newcommand{\fig}[1]{\ref{fig:#1}}

\definecolor{applegreen}{rgb}{0.55, 0.71, 0.0}
\definecolor{byzantine}{rgb}{0.74, 0.2, 0.64}

\newcommand{\cred}[1]{{\color{red}{#1}}}
\newcommand{\cblu}[1]{{\color{blue}{#1}}}

\newcommand{\tbdso}[1]{\cblu{#1}}
\renewcommand{\tbdso}[1]{}

\newcommand{\IntermediateStep}[1]{&\cred{\textrm{\ (($===$ intermediate calc. steps $==>$))}}\nonumber\\ #1  
                                                           &\cred{\textrm{(($<=====================$))}}\nonumber\\}
\renewcommand{\IntermediateStep}[1]{}

\makeatletter
\let\cat@comma@active\@empty
\makeatother

\begin{document}

\preprint{APS/123-QED}

\title{Jones--Roberts solitary waves and the onset of rotation in a spherical surface condensate}

\author{Noel Cuadra}
\affiliation{Kirchhoff-Institut f\"ur Physik, Universit\"at Heidelberg, Im Neuenheimer Feld 227, 69120 Heidelberg, Germany}

\author{Alberto Villois}
\affiliation{School of Engineering, Mathematics and Physics, University of East Anglia, Norwich Research Park, Norwich, NR4 7TJ, United Kingdom}
\affiliation{Centre for Photonics and Quantum Science, University of East Anglia, Norwich Research Park, Norwich, NR4 7TJ, United Kingdom}

\author{Thomas Gasenzer}
\affiliation{Kirchhoff-Institut f\"ur Physik, Universit\"at Heidelberg, Im Neuenheimer Feld 227, 69120 Heidelberg, Germany}
\affiliation{Institut f\"ur Theoretische Physik, Universit\"at Heidelberg, Philosophenweg 16, 69120 Heidelberg, Germany}
\affiliation{ExtreMe Matter Institute EMMI, GSI Helmholtzzentrum f{\"u}r Schwerionenforschung, Planckstrasse 1, 64291 Darmstadt, Germany}

\author{Davide Proment}
\affiliation{School of Engineering, Mathematics and Physics, University of East Anglia, Norwich Research Park, Norwich, NR4 7TJ, United Kingdom}
\affiliation{Centre for Photonics and Quantum Science, University of East Anglia, Norwich Research Park, Norwich, NR4 7TJ, United Kingdom}
\affiliation{ExtreMe Matter Institute EMMI, GSI Helmholtzzentrum f{\"u}r Schwerionenforschung, Planckstrasse 1, 64291 Darmstadt, Germany}


\date{\today}

\begin{abstract}
The nonlinear excitations underlying the onset of rotation in a dilute Bose–Einstein condensate confined to a thin spherical shell are studied.
These excitations correspond to solitary waves rotating about the sphere at constant angular speed: at low speeds they appear as dipoles of singly quantized vortices with opposite circulation, while at higher speeds they evolve into vortex-free Jones--Roberts solitons.
With further increase of the angular speed, these excitations hybridize with equatorially confined modes whose azimuthal wave number is set by the sphere radius measured in units of the healing length.
The propagation speed of these modes is shown to play the role of a Landau critical velocity, thereby setting the upper limiting angular speed of the entire Jones–Roberts family.
\end{abstract}

\maketitle

{\it Introduction.} 
Quantum fluids represent a fascinating field of physics with far-reaching applications that extend well beyond low-temperature physics, impacting areas such as condensed matter, atomic physics, quantum computing, and fundamental quantum phenomena. 
Among quantum fluids, Bose-Einstein condensates (BECs) made of ultracold dilute atomic gases are extremely versatile due to the remarkably precise techniques to control their properties and the well-established first-principle microscopic theory that models them.
For instance, external confining traps can be used to mimic local curvature effects and even attempt to realise (quasi-)two-dimensional quantum fluid films on curved manifolds.
This ultimately opens up the possibility of probing how curvature effects and global topological properties of the confining manifold affect the quantum fluid static and dynamical properties.

Among non-trivial manifolds, the spherical surface is probably the easiest to consider: a compact two-dimensional manifold with constant positive curvature possessing global topological properties different from the Euclidean (flat) plane. 
Specifically, the Euler characteristic of a spherical surface is $\chi_\text{Eul}=2$, while for the plane it is $\chi_\text{Eul}=1$.
Adiabatic trapping potentials have been theoretically proposed for realising spherical shells \cite{zobaygarraway1,zobaygarraway2}, and efforts have been devoted to implement these in experiments \cite{YColombe_2004, Garraway_2016, Guo_2022, Rey2022}.
Various techniques have been considered to counteract thereby the effects of gravity in Earth-based laboratories, from the creation of BECs in Earth-orbiting \cite{Aveline2020, Carollo_2022, BECCAL} or space-born \cite{rocket} microgravity conditions to free-falling BECs dropped within a tower \cite{doi:10.1126/science.1189164, PhysRevLett.129.243402, PhysRevA.101.013634}.
The static properties of a dilute BEC confined in a spherical shell have been extensively studied, including the critical temperature for Bose--Einstein condensation \cite{Tononi_2019, Bereta_2019, Tononi_2020, Rhyno_2021}, the equation of state \cite{Tononi_2022}, and ground state properties for attractive interaction strengths \cite{10.1116/5.0190767}; for an exhaustive review see \cite{Tononi_2024}.

Regarding the dynamical properties of a quantum fluid confined to a thin spherical shell, research has been limited to linear Bogoliubov excitations \cite{Moller_2020}, and semi-analytical and numerical work has been focused on the energetics of a dipole, vortex-antivortex solution \cite{Padavic_2020}, on the emergence of vortex lattices \cite{White_2024}, the statistical features of vortices \cite{PhysRevLett.127.095301, PhysRevResearch.4.013122}, and the effects of rotating the spherical gas in the hydrodynamic framework \cite{PhysRevA.107.023319, doi:10.7566/JPSJ.92.044003}. 
Under the assumption of infinitesimally thin vortex core, the point vortex model has been used to qualitatively model vortex dynamics on spherical surfaces \cite{Bereta_2021}, and other curved two-dimensional manifolds \cite{PhysRevA.96.063608, PhysRevA.99.063602, PhysRevA.105.023307}.

In this Letter, we reveal the fully nonlinear (many-body) excitation family of a dilute quantum fluid confined on the surface of a sphere, containing up to two singly-quantized vortices.
Specifically, we consider a dilute BEC with contact interactions and theoretically model it using the mean-field Gross-Pitaevskii (GP) equation for a thin (quasi-)two-dimensional spherical shell. 
These nonlinear excitations result in solitary wave solutions that move at constant rotational angular speed about any axis of symmetry of the spherical surface, and govern the onset of rotation in the system.
They are the analogue of the axisymmetric solitary waves discovered by Jones \& Roberts \cite{Jones1982MotionsIA} in a quantum fluid confined in a (flat) plane, recently realised in experiment \cite{JRswexp, Baker2025}, but display different properties due to the compactness, curvature, and topological constraints imposed by the spherical surface.
As we shall demonstrate, these excitations depend on the radial size of the system (measured in units of the healing length) and carry lower energy than their linear Bogoliubov counterparts, ultimately unveiling the minimum energy that the system is able to store for a given value of its angular momentum.

\begin{figure*}
\includegraphics[width = \linewidth]{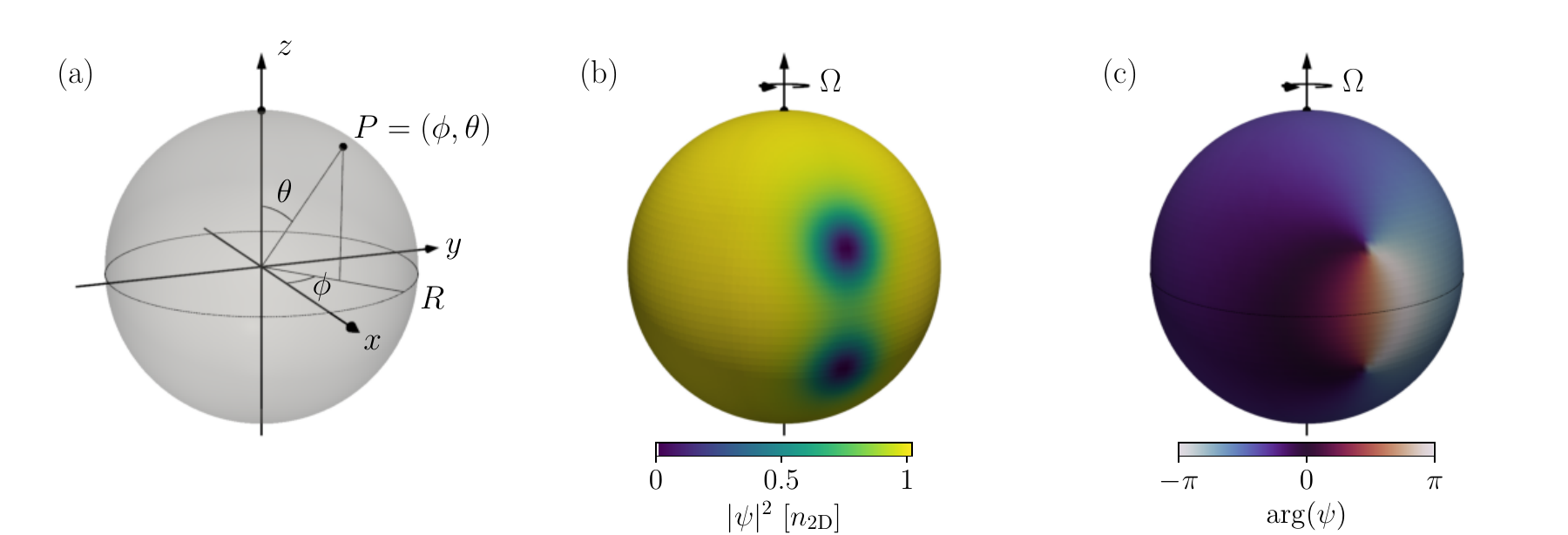}
\caption{\label{fig:SW-example} 
BEC localised within a thin spherical shell of radius $R$. (a) Angular coordinate system $(\phi, \theta)$.
Color-coded representation of (b) the density and (c) the phase angle of the complex field of a solitary wave, consisting of a dipole of two oppositely-charged vortices on a longitude, equidistant from the equator (marked by a thin grey line) and rotating with constant angular speed $\Omega$ about the $z$-axis (here $R = 10$ and $\Omega \approx 0.030 $).
}
\end{figure*}

{\it The GP model on a sphere.}
The dynamics of a (quasi-)two-dimensional BEC confined to the surface of a sphere are modelled by the GP equation,
\begin{equation}
    \i\hbar\frac{\partial\psi}{\partial t} = -\frac{\hbar^2}{2M}\frac{1}{R^2} \Delta_{\phi, \theta}\psi + g_\text{2D}|\psi|^2\psi\, ,
    \label{eq:GP-sph}
\end{equation}
where $M$ is the boson mass, $\Delta_{\phi, \theta} = (\sin\theta)^{-1} \, \partial_\theta (\sin\theta \, \partial_\theta) + (\sin\theta)^{-2} \, \partial_\phi^2$ is the Laplacian over the solid angle, and $g_\text{2D}$ the effective contact interaction on the two-dimensional surface \cite{Tononi_2022}, cf.~\cite{SupplMat} for a derivation. 
The complex order parameter $\psi = \psi(\phi, \theta, t)$ expresses the time evolution of the compact two-dimensional Bose gas with respect to the polar and azimuthal coordinates, $\theta$ and $\phi$, respectively, on a spherical surface of radius $R$, see Fig.~\ref{fig:SW-example}(a).
Upon introducing the mean-field chemical potential $\mu=g_\text{2D}n_\text{2D}$, where $n_\text{2D}=N/(4\pi R^2)$ is the mean two-dimensional density of the $N$ bosons on the sphere, and the healing length $\xi=\sqrt{\hbar^2/(2M\mu)}$, a rescaling of $\psi \to \sqrt{n_\text{2D}}\psi \exp(-\i \mu t)$, $t\to t/\mu$, and $R \to \xi R$ leads to the dimensionless GP equation,
\begin{equation}
    \i \frac{\partial\psi}{\partial t} = -\frac{1}{R^2} \Delta_{\phi, \theta}\psi + |\psi|^2\psi - \psi\, .
    \label{eq:GP-sph-nondim}
\end{equation}
We notice that the dimensionless $R=\sqrt{Mg_\text{2D}N/(2\pi)}$, which measures the radius of the sphere in units of the healing length, is controlled by the coupling and the total boson number and can be effectively used as a free parameter to control the intensity of the kinetic term (first term on r.h.s.) as compared with the nonlinear interaction and average density terms (second and third terms, respectively).

The system conserves the Hamiltonian $H$, the boson number $N$, and, by virtue of the spherical surface symmetry, the angular momentum, which, for the flows we consider, points in $z$-direction, $\mathbf{L}=L_z \vec{e}_z$.
In dimensionless units, they read
\begin{align}
         H &= R^2 \int \left(\frac{1}{R^2} \left|\nabla_{\phi, \theta}\,\psi\right|^2 + \frac{1}{2}|\psi|^4 - |\psi|^2 \right) \text{d}A_{\phi, \theta} \, , \\
         N &= R^2 \int|\psi|^2 \text{d}A_{\phi, \theta} = 1 \, , \\
         L_z &= -\i R^2 \int\psi^\ast\frac{\partial\psi}{\partial\phi} \text{d}A_{\phi, \theta} \, ,
\end{align}
where the integrals are performed over the entire area of the solid angle $A_{\phi, \theta}$, i.e., $\text{d}A_{\phi, \theta}=\sin\theta\, \text{d}\theta\, \text{d}\phi$, $\nabla_{\phi, \theta}  = (\sin\theta)^{-1} \, \partial_\phi {\bf e}_\phi + \partial_\theta {\bf e}_\theta$ is the gradient operator with respect to the solid angle, and the boson number normalisation follows from the dimensionless rescaling. 

The mean-field ground state of the system, obtained by minimising $H$, is given by the homogeneous solution with unitary density $\rho_\text{gs}=|\psi_\text{gs}|^2=1$, therefore evolving as $\psi_{\rm gs}(t)=\exp(-\i t)$, choosing an arbitrary overall constant phase to vanish.
Ground state perturbations $\psi=\{1 + \epsilon [a_{l, m} Y_{l, m}\exp(-\i\omega t) + b^\ast_{l, m} Y_{l, m}^\ast\exp(\i\omega t) \} \psi_{\rm gs}$ in terms of spherical harmonics $Y_{l, m}(\phi, \theta)$ lead, when $0 < \epsilon \ll 1$, to the Bogoliubov-type dispersion relation \cite{Moller_2020},
\begin{equation}
    \omega_{\rm Bog}(l) = \pm \sqrt{ \frac{l(l+1)}{R^2} \left( \frac{l(l+1)}{R^2} + 2 \right) }\, ,
    \label{eq:Bog}
\end{equation}
here stated in its dimensionless form (see \cite{SupplMat} for details).
Note that $\omega_{\rm Bog}(l)$ is independent of the quantum number $m = -l, \dots, l$, as $\Delta_{\phi, \theta} Y_{l, m} = -l(l+1) \ Y_{l, m}$, i.e., the spherical harmonics are eigenfunctions of the Laplacian with eigenvalues only depending on the non-negative integer `orbital' quantum number $l \in \mathbb{N}_0$.

{\it Rotating solitary waves.}
A defining property of a quantum fluid is the quantization of circulation, which implies that rotational flow typically arises through topologically quantized vortex states, since the fluid velocity is proportional to the gradient of the phase of the order parameter, ${\bf v}\propto\nabla\arg(\psi)$.
As a consequence, a finite amount of angular momentum, $L_c$, must be supplied before a state with non-zero global circulation becomes energetically favorable.
In a 2D Euclidean geometry, such states require at least the presence of one or more singly quantized vortex with a non-vanishing net winding number. 
This is due to the Euler characteristic of the (infinite or bounded) plane being $\chi_{\mathrm{Eul}}=1$, and the Poincar{\'e}–Hopf theorem requiring the sum of the Hopf indices of the zeros of any tangent vector field to equal the Euler characteristic. 
A single (anti-)vortex in the plane carries a Hopf index $1$, while the index of a saddle singularity of the current field carries $-1$ inside the plane and $-1/2$ at its edge (which may be at infinity) \cite{PhysRevResearch.7.L022063}.
By contrast, the sphere has $\chi_{\mathrm{Eul}}=2$, so rotational flow can even be realized by, e.g., a pair of quantized vortices with opposite circulation, which, on the sphere, does not lead to a saddle point of the velocity field.
Nevertheless, in spherical geometry, states with angular momentum below $L_c$ may still occur in the form of steadily rotating phase-density disturbances, since the density current ${\bf j}=\rho{\bf v}$ can possess a solenoidal component even in the absence of phase topological defects.
As we show below, these solutions are analogous to Jones–Roberts soliton excitations in the Euclidean case.

Both the vortex-carrying field configurations and the localised phase-density disturbances exist as fully nonlinear solitary wave solutions of the dilute quantum fluid; here we identify such excitations in mean-field approximation given by the GP equation.
By conservation of the angular momentum, we expect them to rotate at constant angular speed $\Omega$, and, as eigenfunctions of the GP Hamiltonian, they thus define the minimum energy required to realize a state of given angular momentum. 
Without loss of generality, we hence choose a coordinate system rotating with $\Omega$ about the $z$-axis, and assume the solution to stationary therein.
Inserting the {\it ansatz} $\psi(\phi, \theta, t) = \psi_{\text{sw}}(\phi-\Omega t, \theta)$ into \Eq{GP-sph-nondim}, we obtain
\begin{equation}
    0 = \i\Omega\frac{\partial\psi_{\text{sw}}}{\partial\phi} - \frac{1}{R^2} \Delta_{\phi, \theta}\psi_{\text{sw}} + |\psi_{\text{sw}}|^2\psi_{\text{sw}} - \psi_{\text{sw}}\, .
\label{eq:SW}
\end{equation}
Given an initial guess sufficiently close to a solution, \Eq{SW} can be solved numerically using an iterative Newton-Raphson (a.k.a.~Newton-Kantorovich) scheme with pseudo-spectral accuracy \cite{boyd2013chebyshev}.
Note that the solitary wave excitation depends on the two continuous parameters $(R, \Omega)$.
Hence, once a solution is found for a specific choice of these, we expect the full family of solutions to be accessible by branch-following continuation, viz.~varying adiabatically $R$ and $\Omega$.

Choosing $R=10$ and a given distance $d_\text{dip}=2\pi R/3$ between the vortex and the antivortex as shown in \Fig{SW-example} we deduce the phase $\varphi_\text{sw}=\arg(\psi_{\text{sw}})$ of the solitary field $\psi_\text{sw}$ by use of the point-vortex model on the surface of a sphere \cite{Newton2001}, which is approximately valid for $R\gg 1$.
We approximate the respective density profiles $\rho_\text{sw}=|\psi_{\text{sw}}|^2$ of the vortices at arc-length distances $d$ (in units of the healing length) from each vortex position as $\rho(d)=(1+2/d^2)^{-1}$, according to an interpolation valid in the flat (Euclidean) two-dimensional GP model \cite{Berloff_2004}. 
With these we obtain an initial guess for the vortex dipole solution and estimate its angular velocity to be $\Omega\approx0.030$ by propagating it with \Eq{GP-sph-nondim}.
We then determine the exact nonlinear solution for the parameters $(R,\Omega)$ from this guess inserted into \Eq{SW} using the Newton-Raphson schere, see \cite{SupplMat} for details.
\Fig{SW-example}(b) and (c) illustrate the resulting density $\rho_\text{sw}$ and phase $\varphi_\text{sw}$, respectively, for $(R,\Omega) \approx (10,0.030)$, two singly-quantized vortices of opposite winding number, with the equator as their perpendicular bisector.
Similarly to the propagation in the Euclidean plane, the dipole's orientation, length, and angular speed $\Omega$ remain constant in time.
See \cite{VideosJRSonSphere} for movies of such dynamics.

By keeping $R=10$ fixed and varying $\Omega$, we then incrementally follow the solution branch.
Similarly to what was previously reported for the Euclidean case \cite{Jones1982MotionsIA}, the dipole size increases with decreasing $\Omega$ and vice-versa.
At a critical, maximum value $\Omega_\text{cr}\approx 0.089$ the two point defects in the phase mutually annihilate, and for larger $\Omega$, the resulting solitary wave solution exhibits a density dip with minimum at the equator, elongated perpendicular to it.
With increasing $\Omega$ its depth diminishes and the transverse elongation grows.
This solution is analogous to the Jones--Roberts soliton in a two-dimensional Euclidean plane \cite{Jones1982MotionsIA}.

\begin{figure}
\includegraphics[width = \linewidth]{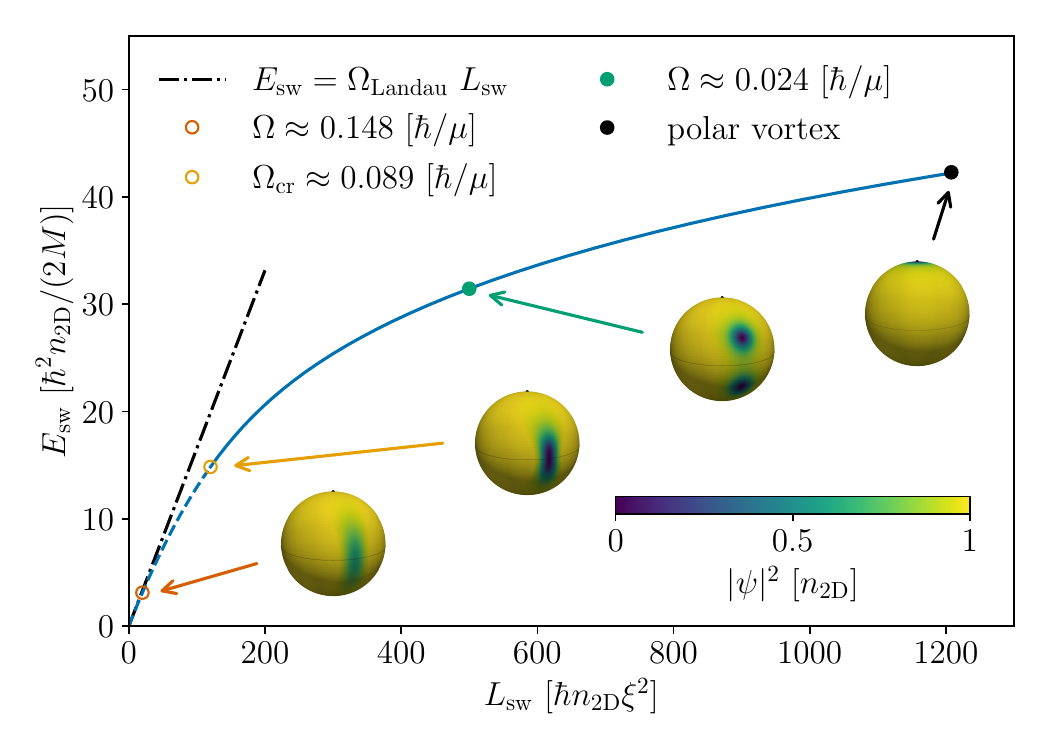}
\caption{\label{fig:E-Lz}
The energies $E_\text{sw}$ of the solitary wave solutions for a given angular momentum $L_\text{sw}$ for the branch with a radius $R=10$ of the sphere in units of the healing length (blue curve); the dotted line indicates solutions above the critical angular velocity $\Omega_\text{cr}$, i.e., vortex-free Jones--Roberts-type solitons. 
The slope of the black dashed-dotted line equals the maximum angular speed $\Omega_\text{Landau}$.
The insets show the density field for four different solutions along the branch: a vortex-free soliton, the solution at the critical $\Omega_\text{cr}$, a dipole, and the polar-vortex configuration where the vortices are sitting at the poles.
}
\end{figure}

It is instructive to compute the energy and angular momentum of the solitary wave excitation, defined as $E_\text{sw} = H[\psi_\text{sw}]-H[\psi_\text{gs}]$ and $L_\text{sw} = L_z[\psi_\text{sw}]$, respectively, noting that $H_\text{gs}=-2 \pi R^2$, $\mathbf{L}_\text{sw} = L_z[\psi_\text{sw}] \mathbf{e}_z$, and $\mathbf{L}[\psi_\text{gs}]=0$.
The energy-angular-momentum dependence of the solitary wave branch for $R=10$ is shown in \Fig{E-Lz}, where the change from continuous to dotted line highlights the transition from vortex dipoles to vortex-free solitons and four examples illustrate the respective densities on the sphere, up to the ``polar-vortex'' configuration, where the two vortices of opposite winding are located at the poles, hence having maximum distance from each other and maximum total vorticity.

Having determined one such branch, we can study how it changes with $R$.
\begin{figure}
\includegraphics[width = \linewidth]{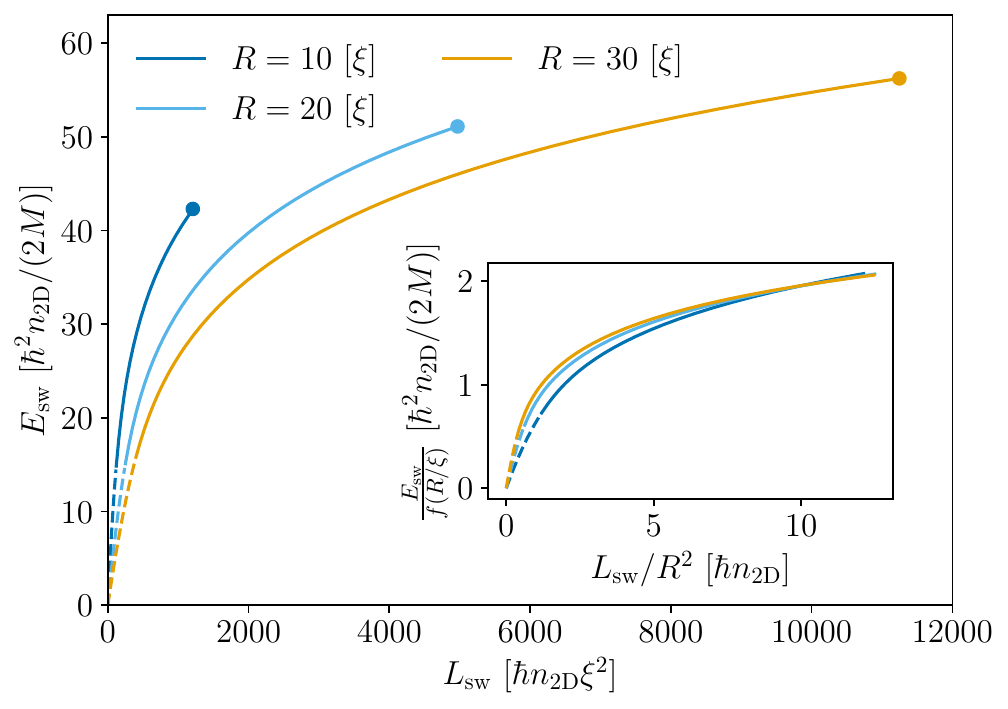}
\caption{\label{fig:E-Lz-many} 
Angular-momentum dependence of the solitary wave energy for different dimensionless radii, $R\in\{10, 20, 30\}$. 
The dotted lines indicate the region of vortex-free density dip solutions and the solid dots mark the polar-vortex solutions.
The inset shows the same branches rescaled according to the Rankine-vortex energy and angular momentum estimates for the polar-vortex solutions where $f(R/\xi) = 1+4\log(2R/\xi)$, see \Eq{scalings}.}
\end{figure}
\Fig{E-Lz-many} shows, in the $(L_\text{sw},E_\text{sw})$-plane, the branches of all possible solutions for $R \in \{10, 20, 30\}$, with the polar-vortex configurations marked by circular dots.
We notice that the maximum angular speed, realized at $L_\text{sw}\to0$, decreases with growing dimensionless radius $R$.
Our results show that the only way to lower the excitation energy of the system at a fixed value of angular momentum, is to increase its radius $R$ relative to the healing length.

Using a Rankine {\it ansatz}, which consists of a patch of solid-body rotation within its core and a zero-vorticity flow outside, we can approximate  $E_\text{sw}$ and $L_\text{sw}$ of the polar-vortex solution, see \cite{SupplMat} for details.
Specifically, we can show that, for $R\gg1$, the angular momentum and energy scale, respectively, as
\begin{equation}
    \begin{split}
        & \lim_{R\gg1} L_\text{sw} \sim R^2 \, , \qquad
          \lim_{R\gg1} E_\text{sw} \sim 1+4\log(2R) \, .
    \end{split}
    \label{eq:scalings}
\end{equation}
The inset of \Fig{E-Lz-many} depicts the three solitary wave branches rescaled accordingly to the Rankine-vortex estimates, highlighting scale invariance being approached for large $R$ for the entire branch.
This is found despite the fact that the Rankine-vortex approximation neither accounts for the correct core properties of a quantum vortex nor matches the vortex-free part of the branch.

We emphasize that, in analogy to Landau's argument for the breakdown of superfluidity, which gives, in the (flat) Euclidean case, Landau's critical fluid velocity $v_\text{Landau} = \min_{p>0} E(p)/p$, in terms of the energy $E(p)$ and linear momentum $p$ of an excitation, we can identify a critical angular velocity $\Omega_\text{Landau}$.
For this, representing the solitary wave excitation branch in the $(L_\text{sw}, E_\text{sw})$-plane is particularly convenient as one can show analytically that the slope of the tangents to the branch equals the respective excitation's angular speed, $\Omega=\partial E_\text{sw}/\partial L_\text{sw}$, see \cite{SupplMat} for a derivation of this identity.

\begin{figure}[ht]
\includegraphics[width=\linewidth]{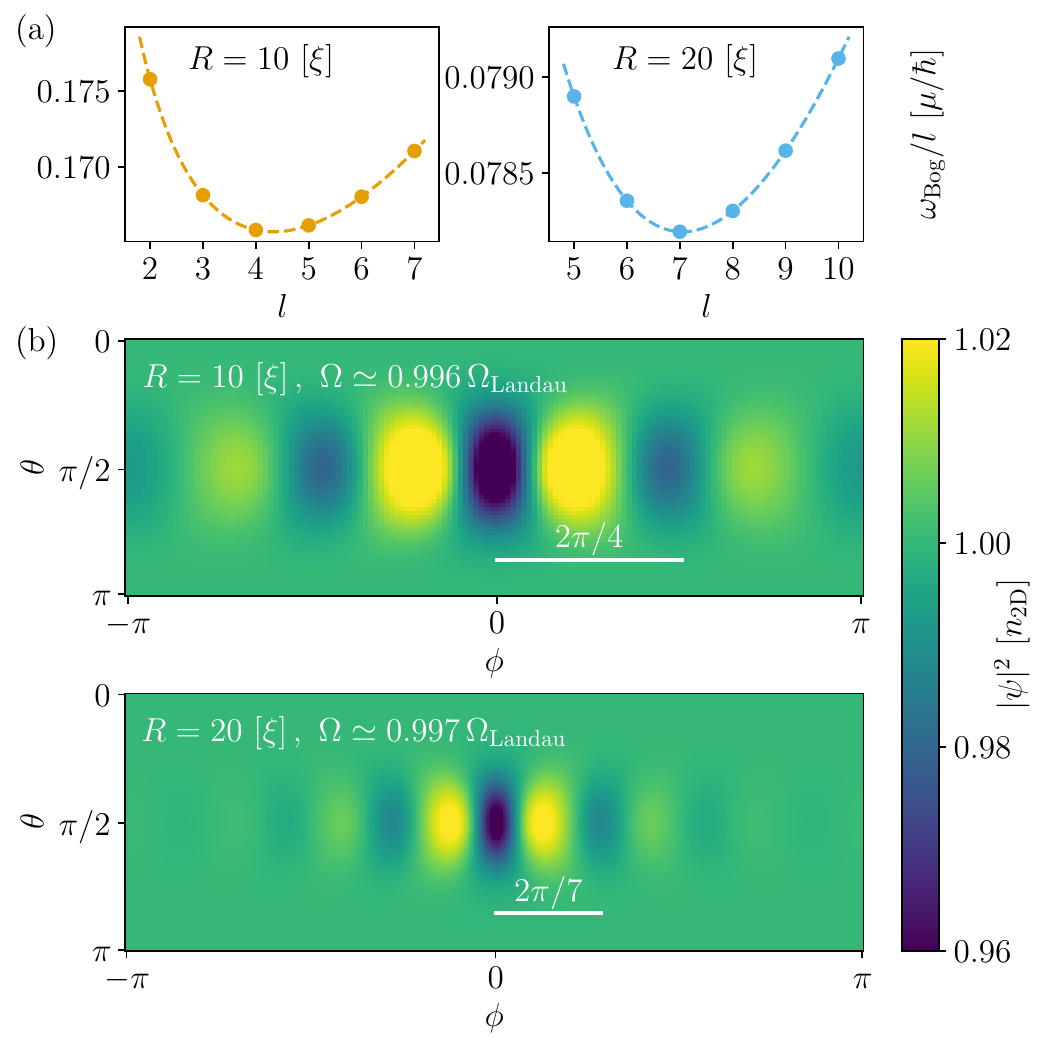}
\caption{\label{fig:omCrit-R}
(a) Fractions $\omega_{\rm Bog}(l)/l$ for Bogoliubov frequencies \eq{Bog}, for two different values of the radius,  $R10$, $20$. 
As the angular-momentum quantum number $l$ is integer-valued, only the dots represent possible angular frequencies, cf.~\Eq{Landau}, while the dashed lines are guides to the eye.
(b) Density fields of the solitary wave solution, in the limit of the Landau critical angular speed $\Omega_\text{Landau}$, for $R=10$ (top) and $R=20$ (bottom).
The white bars indicate the wave angle $2\pi/l_\text{Landau}$ in the $\theta$-direction. 
}
\end{figure}
As seen in Figs.~\fig{E-Lz} and \fig{E-Lz-many}, the solitary wave excitation branches are, for any radius $R$, convex, such that, for each branch, the maximum slope is attained by the vortex-free solution, with energy and angular momentum approaching zero.
In this limit, the solitary wave excitation becomes infinitesimal, thus approaching the Bogoliubov linear excitation, i.e.,
\begin{equation}
    \Omega_\text{Landau} = \min_{L_\text{sw}>0} \frac{E_\text{sw}}{L_\text{sw}} = \min_{l\,\in\,\mathbb{N}} \frac{\omega_{\rm Bog}(l)}{l} \, .
    \label{eq:Landau}
\end{equation}
The second identity was used to determine the slope of the dashed line in \Fig{E-Lz}.
As can be inferred from \Fig{E-Lz-many} and follows from \Eq{Bog}, we notice that $\Omega_\text{Landau}$ depends on $R$, as is explicitly shown in
\Fig{omCrit-R}(a), implying that $\Omega_\text{Landau}=\omega_\text{Bog}(l_\text{Landau})/l_\text{Landau}$, with $l_\text{Landau}(R=10)=4$ and $l_\text{Landau}(R=20)=7$.
We therefore expect the solitary wave solution to exhibit, in the limit of $\Omega \to \Omega_\text{Landau}$, a number of troughs (and peaks) around the sphere corresponding to the value of $l_\text{Landau}$.
\Fig{omCrit-R}(b) confirms this conclusion, showing the solutions for $R=10$ and $R=30$ to possess 4 and 7 troughs (and peaks) in the density, respectively.
Note that, in each case, the amplitude of the density wave decreases away from the central dip, distinguishing the non-linear solitary wave from a delocalised plane-wave Bogoliubov solution.
This unexpected behaviour has no analogue in a Bose gas with contact interactions in a flat (Euclidean) geometry, but it resembles what is found in a dipolar Bose gas where long-range interactions allow for a roton-like minimum in the phase speed of the Bogoliubov excitations, resulting in Landau's critical speed being lower than the bulk sound speed in the limit of vanishing momentum.

We finally analyse this peculiar behaviour in the limit $R \to \infty$, where the positive curvature characteristic of the sphere vanishes and one would naively expect the flat (Euclidean) results to hold.
Interestingly, in the limit $R \to \infty$, $l_\text{Landau}$ scales as $R^{2/3}$ with the sphere's radius $R$ (see \cite{SupplMat} for details on $l_\text{Landau}$). 
Hence, the distance between two neighbouring density minima grows as $2\pi R / l_\text{Landau} \sim R^{1/3}$, which allows recovering the flat (Euclidean) result of a solitary wave \cite{Jones1982MotionsIA} with no disturbances at finite distance.
Moreover, in this limit, the rotational Landau critical speed scales as $\Omega_\text{Landau} \approx \sqrt{2} \ R^{-1}$, hence its radial velocity counterpart, $v_\text{Landau} = \Omega_\text{Landau} R \to \sqrt{2}$, becomes $R$-independent and recovers the Euclidean speed of sound, in non-dimensional units $c=\sqrt{2}$.

{\it Summary and outlook.}
We have demonstrated that rotating solitary wave excitations exist in BECs with contact interactions confined in a thin (quasi-)two-dimensional spherical shell, described by a purely 2D GP defined on the surface of the sphere.
Such solutions form a continuous branch encompassing states ranging from a vortex-antivortex pair at the poles of the sphere, via vortex dipoles with smaller defect distance, moving perpendicular to their polarisation around the sphere, to vortex-free enlongated density dips perpendicular to the equator.
In the limit of vanishing angular momentum and energy they form localised, finite-width Bogoliubov-like density waves propagating along the equator.
Our findings constitute the analogue of the Jones--Roberts solitary wave branch obtained for (quasi-)2D Bose gases in a flat Euclidean geometry.

Beyond our focus being set on Bose gases confined within thin spherical shells, the results outlined in this Letter open a wider perspective on adjacent research areas.
In applied mathematics, specifically in the context of nonlinear waves, they prompt further studies concerning the emergence of solitary waves in curved geometries, as well as their stability \cite{CAJones_1986, PhysRevA.105.063325, Krause2024a.PhysRevA.110.053302}. 
We can show that two such solitary waves with equal and opposite rotation speed, being initially placed sufficiently far apart survive several scattering processes, see \cite{VideosJRSonSphere} for a movie of such dynamics.
It would furthermore be interesting to search for similarities with driven-dissipative situations, such as solitons arising in Kerr optical whispering-gallery-mode resonators being modelled by the Lugiato–Lefever equation \cite{PhysRevA.87.053852, Herr2014}.
Finally, in low-temperature physics, they hint at possible similar excitations in multielectron bubbles in liquid helium \cite{PhysRevB.65.195418, PhysRevB.72.094506, PhysRevB.74.104512}, in superconducting spherical shells \cite{Tempere2007, PhysRevB.79.134516}, in the superfluid-helium coating of spheres \cite{Wang_1986}, and in three-dimensional superfluids confined in spherical geometries ranging from liquid-helium drops \cite{PhysRevLett.130.216001}, via BEC spherical shells of finite thickness \cite{Lannert_2007, Sun_2018, Padavic_2017}, quantum droplets in BEC mixtures \cite{Ruban_2022, PhysRevLett.134.093401, dbm1-7dy4}, to vortex configurations in rotating BEC bubbles \cite{PhysRevA.109.013301}.

\emph{Acknowledgments}.
The authors thank Gregor Bals, Carlo Ewerz, John Hannay, Georg Trautmann, and Martin Zboron for discussions and collaboration on related topics.
The authors acknowledge support by EPSRC, Grant Number EP/R014604/1, by the ExtreMe Matter Institute EMMI at the GSI Helmholtzzentrum f{\"u}r Schwerionenphysik, Darmstadt, Germany, by the Deutsche Forschungsgemeinschaft (DFG, German Research Foundation), through SFB 1225 ISOQUANT (Project-ID 273811115), grant GA677/10-1, and under Germany's Excellence Strategy -- EXC 2181/1 -- 390900948 (the Heidelberg STRUCTURES Excellence Cluster), and by the state of Baden-W{\"u}rttemberg through bwHPC, the data storage service SDS{@}hd supported by the Ministry of Science, Research and the Arts Baden-W{\"u}rttemberg (MWK), and the DFG through grants INST 35/1503-1 FUGG, INST 35/1597-1 FUGG, and INST 40/575-1 FUGG (SDS, Helix, and JUSTUS 2).

\begin{appendix}
\begin{center}
\textbf{APPENDIX}
\end{center}
\setcounter{equation}{0}
\setcounter{table}{0}
\makeatletter

In the appendix, we provide details of the formulation of the GP model on the spherical surface, its linear excitations, the numerical methods chosen, the Rankine polar-vortex solution, the Landau critical angular speed, and the spherical harmonics components of the solitary wave solution.

\section{The thin spherical shell confinement about \texorpdfstring{$r=R$}{r = R}}
The Gross-Pitaevskii equation for a Bose--Einstein condensate reads
\begin{equation}
    \i\hbar\frac{\partial\Psi}{\partial t} = - \frac{\hbar^2}{2M}\nabla^2 \Psi + V_\text{ext}\Psi + \frac{4\pi\hbar^2 a_{\rm s}}{M}|\Psi|^2\Psi \, ,
    \label{eq:SM:GP-3D}
\end{equation}
where $M$ and $a_{\rm s}$ are the mass and the s-wave scattering length of the boson, respectively, and $V_{\rm ext}$ is a confining external potential.

We assume the confining potential spherically symmetric, having a global minimum about $r=R$, and strong enough to be able to freeze any dynamics along the radial direction.
It is therefore convenient to go from Cartesian coordinates $(x, y, z)$  to spherical coordinates $(r, \phi, \theta)$ according to
\begin{align}
    & x=r\sin\theta\cos\phi \, , \\
    & y=r\sin\theta\sin\phi \, , \\
    & z=r\cos\theta\, ,
\end{align}
with $\phi \in [0, 2\pi]$ and $\theta \in [0, \pi]$.
We impose the ansatz
\begin{equation}
    \Psi(r, \phi, \theta, t)=\psi(\phi, \theta, t) h(r) \exp(-\i\mu_\text{ext}t/\hbar) \, ,
    \label{eq:SM:ansatzConf}
\end{equation}
where $\mu_\text{ext}$ is the added chemical potential energy due to the external confinement.
The form of $h(r)$ is such to cancel out the external potential term in \Eq{SM:GP-3D}; for instance, a Gaussian corresponding to the ground state eigenfunction of the harmonic oscillator cancels out a harmonic potential of the form $V(r) \propto (R-r)^2$.

After substituting \Eq{SM:ansatzConf} into \Eq{SM:GP-3D}, expressing the Laplacian operator in spherical coordinates, and integrating along the radial direction, we finally obtain
\begin{equation}
    \i\hbar\frac{\partial\psi}{\partial t} = -\frac{\hbar^2}{2M}\frac{1}{R^2} \Delta_{\phi, \theta}\psi + g_\text{2D}|\psi|^2\psi\, ,
    \label{eq:SM:GP-sph}
\end{equation}
where $g_\text{2D} $ is an effective two-dimensional coupling coefficient.

\section{Bogoliubov excitations}
The ground state solution of \Eq{SM:GP-sph} corresponds to the homogeneous solution
\begin{equation}
    \psi_\text{gs}(t) = \sqrt{\rho_{\rm gs}} \, \e^{-\i(\mu t/\hbar + \nu)} \, , \quad \mu=\rho_\text{gs} g_\text{2D} \, ,
    \label{eq:SM:unifSol}
\end{equation}
where $\nu \in [0, 2\pi)$ is an arbitrary initial phase that can be set to zero without loss of generality.
It is instructive to characterise the infinitesimal perturbation $p(\phi, \theta, t)$ of the ground state; this is done by setting $0 < \epsilon \ll 1$, substituting the ansatz
\begin{equation}
    \psi(\phi, \theta, t) = \left[1 + \epsilon p(\phi, \theta, t) \right] \sqrt{\rho_{\rm gs}} \, \e^{-\i\mu t/\hbar}
\end{equation}
into \Eq{SM:GP-sph}, and solve order by order in $\epsilon$.

At order $\epsilon^0$, the equation is automatically satisfied as this corresponds simply to the ground state solution.
At order $\epsilon^1$, the equation results in the following Schr{\"o}dinger-like equation for the perturbation
\begin{equation}
    \i\hbar \frac{\partial p}{\partial t} = -\frac{\hbar^2}{2M} \frac{1}{R^2}\Delta_{\phi, \theta}p +\mu(p + p^\ast) \, .
    \label{eq:SM:GPSS-BogEps}
\end{equation}
In the Bogoliubov spirit, we assume that the perturbation is the following superposition of spherical harmonics
\begin{equation}
    p(\phi, \theta, t) = a_{l, m} Y_{l, m}(\phi, \theta)\e^{-\i \omega t} + b_{l, m}^\ast Y_{l, -m}(\phi, \theta)\e^{\i \omega t}  \, ,
    \label{eq:SM:GPSS-BogAnsatz}
\end{equation}
where we have used the identity $ Y_{l, m}^\ast(\phi, \theta) = (-1)^m \, Y_{l, -m}(\phi, \theta) $.
By substituting \Eq{SM:GPSS-BogAnsatz} into \Eq{SM:GPSS-BogEps} and setting both $ Y_{l, m} $ and $ Y_{l, -m} $ coefficients to zero, we find
\begin{equation}
    \left\{
    \begin{aligned}
     a_{l, m}\left[ -\hbar\omega + \frac{\hbar}{2M}\frac{l(l+1)}{R^2} + \mu \right] + b_{l, m}\left(\mu \right) 
     &= 0 \\
     a_{l, m}^\ast\left(\mu \right) + b_{l, m}^\ast\left[ \hbar\omega + \frac{\hbar}{2M}\frac{l(l+1)}{R^2} + \mu \right] 
     &= 0
    \end{aligned}
    \right.
\end{equation}
The solvability condition gives the Bogoliubov-type dispersion relation
\begin{equation}
    \hbar^2\omega_\text{Bog}^2(l) = \left[\frac{\hbar^2}{2M}\frac{l(l+1)}{R^2}\right]^2 +2\mu\left[\frac{\hbar^2}{2M}\frac{l(l+1)}{R^2}\right]  \, ,
\end{equation}
which further simplifies to
\begin{equation}
    \omega_\text{Bog}(l) = \pm\sqrt{\frac{\hbar}{2M}\frac{l(l+1)}{R^2}} \sqrt{\frac{\hbar}{2M}\frac{l(l+1)}{R^2}+\frac{2\mu}{\hbar}}  \, .
    \label{eq:SM:BogDisp}
\end{equation}

\section{Numerical methods}
\subsection{Initial condition}
Our numerical analysis begins with the initial condition for a pair of vortices on the surface of the sphere. In the flat plane, the phase profile of a vortex-antivortex pair, situated at positions $(x_+, y_+)$ and $(x_-,y_-)$, takes the well-known form 
\begin{equation}
    S_{\text{pl}}(x, y) = \text{atan2}(y - y_+, x - x_+)  - \text{atan2}(y - y_-, x - x_-)\,,
\end{equation}
where $\text{atan2}$ is the two-argument arctangent. This phase is mapped onto the spherical surface using the stereographic projection
\begin{equation}
    (x, y) = \cot{\frac{\theta}{2}} (\sin{\phi}, \cos{\phi})\,.
\end{equation}
Assuming that the dipole sits on the antimeridian, the vortex is on the Northern Hemisphere, and the antivortex on the Southern Hemisphere, the projection yields
\begin{align}
        S(\phi, \theta) &= \text{atan2} \left( \cot{\frac{\theta}{2}} \cos{\phi} + \cot{\frac{\theta_+}{2}}, \  \cot{\frac{\theta}{2}}\sin{\phi} \right) \nonumber \\
    &- \text{atan2} \left(\cot{\frac{\theta}{2}} \cos{\phi} + \tan{\frac{\theta_+}{2}}, \ \cot{\frac{\theta}{2}}\sin{\phi} \right)\,,
    \label{eq:phase}
\end{align}
where the phase is now solely determined by the polar angle of the vortex, $\theta_+ \in \left [ 0, \frac{\pi}{2} \right)$. Note that the stereographic projection is a conformal mapping, i.e. it preserves angles, from which it follows that circles on the plane are mapped onto circles on the sphere. However, it does not preserve either lengths or areas. It is therefore expected that \Eq{phase} introduces distortions into the phase profile, yet it is not critical to get the phase exactly right. The initial condition proposed can be subject to a short imaginary-time evolution to properly form the vortex cores and smooth out any distortions.

\subsection{Spectral methods}
The order parameter $\psi$ is represented using spherical harmonics $Y_{l, m}(\theta,\phi)$ with complex coefficients $c_{l, m}$,
\begin{equation}
    \psi(\phi, \theta) = \sum_{l=0}^{l_{\max}}\sum_{m=-l}^l c_{l, m} Y_{l, m}(\phi, \theta) \,,
\end{equation}
whose indices are truncated to $0 \le l \le l_{\max}$ and $|m| \le l_{\max}$ for numerical purposes.
Two different numerical libraries have been used in the numerical simulations, both leading to the same results up to spectral accuracy: we made use of the SHTOOLS archive \cite{SHTools} and of the Python wrapper of the SHTns libraries \cite{https://doi.org/10.1002/ggge.20071}.

When using the SHTools archive, the spherical coordinates $\theta$ and $\phi$ that parametrize the sphere are mapped onto an equally spaced $N \times 2N$ grid. It follows that the angle between two neighboring grid points in both directions is $\pi/N$ and the angles of each point are given by
\begin{align}
    \theta_i = \frac{i \pi}{N}, \ \phi_j = \frac{j \pi}{N}\,, 
\end{align}
where $i \in \{0, 1, \ldots, N - 1\}$ and $j \in \{0, 1, \ldots, 2N - 1\}$. The order parameter $\psi$ is then formally stored as a 2D array of shape $(N, 2N)$ and the entry with index $[i,j]$ gives the complex number $\psi(\theta_i,\phi_j)$, i.e., the rows of the array represent constant latitude and the columns constant longitude. The bands at the south pole ($\theta = \pi$) and at $\phi = 2\pi$ (which is the same as $\phi = 0$) are not required by the transformation routines and are thus not included in the grid data.
The expansion coefficients $ c_{l, m} $ are then calculated using the sampling theorem by Driscoll and Healy \cite{DRISCOLL1994202}.
The cutoff $l_{\max}$ is limited by the grid size to $l_{\text{max}} = N/2 - 1$. 

When using the SHTns libraries, one has the freedom to work with different collocation grids.
A standard choice is choosing a non-uniform Chebyshev-like discretisation along the $\theta$-axis and a uniform grid along the $\phi$-axis; the number of collocation points is then automatically chosen by the SHTns library when fixing $l_{\max}$ and $m_{\max} = \max |m|$.
For example, the choice $l_{\max}=m_{\max}=64$ leads to $N_\theta = 66$ and $N_\phi= 140$ points, respectively.

Independently of the numerical library chosen, the spherical-harmonics decomposition lets us calculate the differential operators in \Eq{SW} with spectral accuracy. 
Since the spherical harmonics are eigenfunctions of $\Delta_{\phi, \theta}$, with eigenvalues $-l(l+1)$, the action of the Laplacian $\Delta_{\phi, \theta}$ on the order parameter results in
\begin{align}
    \Delta_{\phi, \theta} \psi(\phi, \theta) = - \sum_{l=0}^{l_{\max}}\sum_{m=-l}^l l(l+1) \, c_{l, m} Y_{l, m}(\phi, \theta)\,.
\end{align}
Similarly, applying $\partial_\phi$ to the order parameter results in 
\begin{align}
    \partial_\phi \psi(\phi, \theta) = \i \sum_{l=0}^{l_{\max}}\sum_{m=-l}^l m \, c_{l, m} Y_{l, m}(\phi, \theta)\,.
\end{align}
%

\subsection{Newton-Raphson method}
Our goal is to find a solution of 
\begin{equation}
    0 = \i\Omega\frac{\partial\psi_{\text{sw}}}{\partial\phi} - \frac{1}{R^2} \Delta_{\phi, \theta}\psi_{\text{sw}} + |\psi_{\text{sw}}|^2\psi_{\text{sw}} - \psi_{\text{sw}}\, .
\label{eq:GP-SW}
\end{equation}
In the following we drop the index SW for simplicity.
We define the complex functional $F$ for the Newton-Raphson method using the previous equation, resulting in
\begin{equation}
    F(\Re\psi, \Im\psi) = \i\Omega\frac{\partial\psi}{\partial\phi} - \frac{1}{R^2} \Delta_{\phi, \theta}\psi + |\psi|^2\psi - \psi \, .
    \label{eq:F}
\end{equation}
Given the root of the functional $F$ being $f_{\rm{root}} = f_{\rm{guess}} - \delta f$, where $f=(\Re\psi, \Im\psi)^T$ is formally an array storing the dependence on the unknown field $\psi$, the Newton-Raphson method iteratively solves the linear equation
\begin{equation}
    A \, \delta f = b\,,
    \label{eq:LAP}
\end{equation}
where
\begin{widetext}  
\begin{align}
    A= \left.
    \begin{pmatrix}
        \frac{\delta \Re F}{\delta \Re\psi} & \frac{\delta \Re F}{\delta \Im\psi} 
        \\[1.5ex]
        \frac{\delta \Im F}{\delta \Re\psi} & \frac{\delta \Im F}{\delta \Im\psi}
    \end{pmatrix} \right|_{f_{\rm{guess}}} 
    &= \left.
        \begin{pmatrix}
            - \frac{1}{R^2} \Delta_{\phi, \theta} - 1 + \left(3\Re\psi^2 + \Im\psi^2\right) & -\Omega\frac{\partial}{\partial\phi} - 2\Re\psi\Im\psi 
            \\[1.5ex]
            \Omega\frac{\partial}{\partial\phi} - 2\Re\psi\Im\psi & - \frac{1}{R^2} \Delta_{\phi, \theta} - 1 + \left(\Re\psi^2 + 3\Im\psi^2\right)
        \end{pmatrix} \right|_{f_{\rm{guess}}}\,,
    \\
    b= \left.
        \begin{pmatrix}
            \Re F \\[1ex]
            \Im F 
        \end{pmatrix} \right|_{f_{\rm{guess}}} 
    &= \left.
        \begin{pmatrix}
            -\Omega\frac{\partial\Im\psi}{\partial\phi} - \frac{1}{R^2} \Delta_{\phi, \theta}\Re\psi + \left(\Re\psi^2 + \Im\psi^2\right)\Re\psi - \Re\psi 
            \\[1ex]
            \Omega\frac{\partial\Re\psi}{\partial\phi} - \frac{1}{R^2} \Delta_{\phi, \theta}\Im\psi + \left(\Re\psi^2 + \Im\psi^2\right)\Im\psi - \Im\psi 
        \end{pmatrix} \right|_{f_{\rm{guess}}} \, .
\end{align}
\end{widetext}
The linear algebra problem is solved by employing pseudo-spectral techniques, namely computing the differential operators by means of spherical harmonics decomposition, and using a generalised minimal residual method (LGMRES) \cite{LGMRES}.

\section{Variational approach and rotation frequency}
Using the dimensionless formulation, the Hamiltonian and the $z$-component of the angular momentum of the system are
\begin{equation}
    H = R^2 \int \left(\frac{1}{R^2} \left|\nabla_{\phi, \theta}\psi\right|^2 + \frac{1}{2}|\psi|^4 - |\psi|^2 \right) \text{d}A_{\phi, \theta}
\end{equation}
and
\begin{equation}
    L_z = -\i R^2 \int\psi^\ast\frac{\partial\psi}{\partial\phi} \text{d}A_{\phi, \theta} \, ,
\end{equation}
respectively.
As for any excitation of the ground state, the energy of the the solitary wave is given by the solitary wave Hamiltonian minus the ground state Hamiltonian, resulting in
\begin{equation}
    \begin{split}
        E_{\rm sw} 
        &= H(\psi_{\rm sw}, \psi_{\rm sw}^\ast) - H(\psi_{\rm gs}, \psi_{\rm gs}^\ast) \\
        & = R^2 \int \left[\frac{1}{R^2} \left|\nabla_{\phi, \theta}\psi_{\rm sw}\right|^2 + \frac{1}{2}\left(|\psi_{\rm sw}|^2 - 1\right)^2 \right] \text{d}A_{\phi, \theta} \, ,
    \end{split}
\end{equation}
as $|\psi_{\rm gs}|^2 = 1$.
We now show how the solitary wave energy $E_{\rm sw}$ varies with its angular momentum.
Note that, by construction, the angular momentum of the solitary wave has a non-zero component only along the $z$-axis and the angular momentum of the ground state vanishes. 
Hence, the solitary-wave angular momentum is $L_{\rm sw} = L_{\rm z}(\psi_{\rm sw}, \psi_{\rm sw}^\ast)$.
We thus find that
\begin{equation}
    \frac{\partial E_{\rm sw}}{\partial L_{\rm sw}} 
    = \frac{\delta H_{\rm sw}}{\delta L_{\rm sw}} \, ,
\end{equation}
where the energy and angular momentum variations are calculated using functional derivatives.
As the ground state Hamiltonian is constant, the energy variation of the solitary wave results as
\begin{widetext}
\begin{equation}
    \begin{split}
        \delta E_{\rm sw} = & \left.\frac{\delta H}{\delta \psi}\right|_{\psi_{\rm sw}, \psi_{\rm sw}^\ast} \delta\psi + \left.\frac{\delta H}{\delta \psi^\ast}\right|_{\psi_{\rm sw}, \psi_{\rm sw}^\ast} \delta\psi^\ast 
        =  R^2 \int \left[\frac{1}{R^2} (\nabla_{\phi, \theta}\psi_{\rm sw}^\ast)\cdot(\nabla_{\phi, \theta}\delta\psi) + |\psi_{\rm sw}|^2\psi_{\rm sw}^\ast\delta\psi - \psi_{\rm sw}^\ast\delta\psi + \text{c.c.} \right] \text{d}A_{\phi, \theta} \, ,
    \end{split}
    \label{eq:Envar}
\end{equation}
\end{widetext}
where $(\delta\psi, \delta\psi^\ast)$ represent a generic infinitesimal variation, and c.c.~denotes the complex conjugate.
The angular momentum variation of the solitary wave reads

\begin{equation}
    \begin{split}
        \delta L_{\rm sw} & = \left.\frac{\delta Lz}{\delta \psi}\right|_{\psi_{\rm sw}, \psi_{\rm sw}^\ast} \delta\psi + \left.\frac{\delta L_z}{\delta \psi^\ast}\right|_{\psi_{\rm sw}, \psi_{\rm sw}^\ast} \delta\psi^\ast \\
        & = -\i R^2 \int \left[ \psi_{\rm sw}^\ast\frac{\partial}{\partial \phi}\left(\delta\psi\right) + \text{c.c.} \right] \text{d}A_{\phi, \theta} \, .
    \end{split}
\end{equation}
Note that the first term in the energy variation can be recast by integration by parts and using the complex conjugate of the solitary wave \Eq{GP-SW}, giving
\begin{equation}
    \begin{split}
        & \int \frac{1}{R^2} (\nabla_{\phi, \theta}\psi_{\rm sw}^\ast)\cdot(\nabla_{\phi, \theta}\delta\psi) \text{d}A_{\phi, \theta} \\
        & = -\int \frac{1}{R^2} \delta\psi \Delta_{\phi, \theta}\psi_{\rm sw}^\ast \text{d}A_{\phi, \theta} \\
        & = \int \delta\psi \left( \i\Omega\frac{\partial\psi_{\text{sw}}}{\partial\phi} - |\psi_{\text{sw}}|^2\psi_{\text{sw}} + \psi_{\text{sw}}    \right) \text{d}A_{\phi, \theta} \, .
    \end{split}
\end{equation}
With this, the energy variation \eq{Envar} can be simplified as follows,
\begin{equation}
    \begin{split}
        \delta E_{\rm sw} & = \i\Omega R^2 \int \left[ \delta\psi\frac{\partial\psi_{\rm sw}^\ast}{\partial \phi} + \text{c.c.}  \right] \text{d}A_{\phi, \theta} \\
        & = -\i\Omega R^2 \int \left[ \psi_{\rm sw}^\ast \frac{\partial}{\partial \phi}\left(\delta\psi\right) + \text{c.c.}  \right] \text{d}A_{\phi, \theta} \\
        & = \Omega\, \delta L_{\rm sw} \, ,
    \end{split}
\end{equation}
which proves that
\begin{equation}
    \frac{\partial E_{\rm sw}}{\partial L_{\rm sw}} = \frac{\delta E_{\rm sw}}{\delta L_{\rm sw}} = \Omega \, .
\end{equation}
%

\section{The Rankine polar vortex solution}
In the flat (Euclidean) plane, the Rankine vortex is a rotational velocity field about the origin, where the motion inside the vortex core is equivalent to a solid-body rotation, while outside it is represented by an irrotational flow; the two flows match at the vortex core boundary ensuring continuity of the velocity field.
We use an analogue flow on the sphere, where the velocity flow rotates about the $z$-axis, is non-zero along the azimuthal, $\phi$, direction, and is independent of the $\phi$ coordinate.
Representing the core size as an angle $\theta_\xi$, the Rankine polar-vortex solution on the sphere is defined as
\begin{equation}
    v_\phi(\theta)= \left\{
    \begin{aligned}
        & \frac{\Gamma}{2\pi R \sin\theta_\xi}R \sin\theta \, , \quad \theta\le\theta_\xi \\
        & \frac{\Gamma}{2\pi R \sin\theta} \, , \quad \theta_\xi<\theta<\pi-\theta_\xi \\
        & \frac{\Gamma}{2\pi R \sin\theta_\xi}R \, , \quad \theta\ge\pi-\theta_\xi \\
    \end{aligned}
    \right. \, ,
\end{equation}
where $\Gamma$ is the circulation of the vortex at the North pole.

By construction and symmetry, the total angular momentum of the system corresponds to its $z$-component and results in
\begin{equation}
    L_z=\rho \int \left(R \sin\theta\right) v_\theta R^2 \text{d}A_{\phi, \theta} \, ,
\end{equation}
where $\rho$ is the (constant) density of the fluid.
The fluid's energy is simply given by the kinetic energy, resulting in
\begin{equation}
    E_\text{kin}=\frac{\rho}{2} \int v_\theta^2 R^2 \text{d}A_{\phi, \theta} \, .
\end{equation}
These two integrals can be evaluated in closed form and weakly depend on the specific choice of the core size $\theta_\xi$, see \cite{NotebooksJRSonSphere} for more details.
In the following we choose $\theta_\xi = 1/R$, i.e., a core arc-length radius of $R\theta_\xi = 1$, in units of the healing length. 

By taking the limit $R\gg1$, where one expects the Rankine vortex solution to work better, we can show that
\begin{equation}
    \lim_{R\to\infty} L_z = 2\rho\Gamma R^2 + \mathcal{O}\left(1\right)
\end{equation}
and
\begin{equation}
    \lim_{R\to\infty} E_\text{kin} = \frac{\rho\Gamma^2\left[1+4\log\left(2R\right)\right]}{8\pi} + \mathcal{O}\left({R}^{-2}\right) \, .
\end{equation}
Hence, we find that the angular momentum scales as $R^2$, while the energy of the fluid scales as a non-trivial function $f$ of the radius, namely
\begin{equation}
    f(R) = 1+4\log\left(2R\right) \, .
\end{equation}
%

\section{The Landau critical angular speed}
The dimensionless Bogoliubov dispersion relation reads
\begin{equation}
    \omega_{\rm Bog} = \sqrt{\frac{l(l+1)}{R^2} \left( \frac{l(l+1)}{R^2} + 2 \right)} \, .
\end{equation}
Considering $l$ a continuous parameter instead of being discrete, we find the extrema of the fraction $\omega_{\rm Bog}(l)/l$ by differentiating by $l$ and setting it to zero.
This results in
\begin{equation}
    \frac{d}{dl} \left( \frac{\omega_{\rm Bog}}{l} \right) = \frac{R}{l} \frac{l^2+l^3-R^2}{\sqrt{l(l+1)\left( l+l^2+2R^2 \right)}} = 0 \, ,
\end{equation}
and therefore the extrema are situated at the roots of
\begin{equation}
    l^2+l^3-R^2 = 0 \, .
\end{equation}
This third-order polynomial in $l$, for $R \in \mathbb{R}$ has only one real solution, resulting in (see \cite{NotebooksJRSonSphere} for details)
\begin{widetext}
\begin{equation}
    l^\star = \frac{1}{3} \left\{ -1 + \frac{1}{\left[ -1 + \frac{3 R}{2} \left( 9R + \sqrt{81 R^2 - 12} \right) \right]^{1/3}} + \left[ -1 + \frac{3 R}{2} \left( 9R + \sqrt{81 R^2 - 12} \right) \right]^{1/3} \right\} \, .
\end{equation}
\end{widetext}
Therefore the value of $l$ referring to the Landau critical speed is either the floor or the ceiling of $l^\star$, whichever gives the minimal value of $\omega_{\rm Bog}(l)/l$.

In the limit $R \to \infty$, this expression can be expanded in powers of $R$, 
\begin{equation}
    \lim_{R \to \infty} l^\star = R^{2/3} + \mathcal{O}(1) \, . 
\end{equation}
%

\section{Spherical harmonics components of the solitary wave solution}
When decomposing the solitary wave solution $\psi_{\rm sw}(\phi-\Omega t, \theta)$ in spherical harmonics, we find
\begin{equation}
    \begin{split}
     \psi_{\rm sw}(\phi-\Omega t, \theta) 
    & = \sum_{l=0}^\infty \sum_{m=-l}^{m} c_{l, m} Y_{l, m}(\phi-\Omega t, \theta) \\
    & = \sum_{l=0}^\infty \sum_{m=-l}^{m} c_{l, m} \e^{\i m (\phi-\Omega t)} P_{l, m}(\cos\theta) \\
    & = \sum_{l=0}^\infty \sum_{m=-l}^{m} d_{l, m}(t) \e^{\i m \phi} P_{l, m}(\cos\theta) \, ,
    \end{split}
\end{equation}
where $P_{l, m}(\cdot)$ are the associated Legendre polynomials of index $(l, m)$, and we have defined the time-dependent spherical harmonics coefficients
\begin{equation}e
    d_{l, m}(t) = c_{l, m} \e^{-\i \Omega t} \, .
\end{equation}
Assuming that these coefficients are periodic in time, we can compute their inverse Fourier transform,
\begin{equation}
    \begin{split}
        \tilde{d}_{l, m}(\omega') & = \frac{1}{2\pi} \int d_{l, m}(t) \e^{\i \omega' t} dt \\
        & = d_{l, m} \delta_{\rm D}(\omega'-m\Omega) \, ,
    \end{split}
\end{equation}
where $\delta_{\rm D}(\cdot)$ is Dirac's $\delta$-function.
This results means that, for any given $l$, the accessible azimuthal quantum numbers $m = -l, \dots, l$ give a resonance at $\omega' = m \Omega$.

To avoid resonant interactions between these solitary wave solutions and the infinitesimal Bogoliubov modes, we must have that
\begin{equation}
    \omega' \neq \omega_{\rm Bog}(l) \quad \Longleftrightarrow \quad \Omega l < \omega_{\rm Bog}(l) \, , \ \forall \ l \in \mathbb{N} \, .
\end{equation}
By dividing the last inequality by $l$, and recalling the definition of Landau's critical angular speed, \Eq{Landau}, we find that
\begin{equation}
    \Omega < \Omega_\text{Landau} \, ,
\end{equation}
ultimately confirming that, in order not to radiate Bogoliubov modes, the solitary wave solution must rotate at an angular speed lower than Landau's critical one.

\end{appendix}

\providecommand{\noopsort}[1]{}\providecommand{\singleletter}[1]{#1}%

\end{document}